\documentclass[oneside,english]{elsart}
\usepackage[T1]{fontenc}
\usepackage[latin1]{inputenc}
\usepackage{graphicx}
\usepackage{setspace}
\usepackage{amssymb}

\makeatletter

\usepackage{babel}
\makeatother
\begin{document}
\begin{frontmatter}

\title{The Stonehenge Technique. A new method for aligning coherent bremsstrahlung
radiators.}

\author{Ken Livingston}

\address{Department of Physics \& Astronomy, University of Glasgow, Glasgow,
G12 8QQ}

\ead{k.livingston@physics.gla.ac.uk}

\begin{abstract}
This paper describes a new technique for the alignment of crystal
radiators used to produce high energy, linearly polarized photons
via coherent bremsstrahlung scattering at electron beam facilities.
In these experiments the crystal is mounted on a goniometer which
is used to adjust its orientation relative to the electron beam. The
angles and equations which relate the crystal lattice, goniometer
and electron beam direction are presented here, and the method of
alignment is illustrated with data taken at MAMI (the Mainz microtron).
A practical guide to setting up a coherent bremsstrahlung facility
and installing new crystals using this technique is also included.
\end{abstract}
\end{frontmatter}

\section{Introduction\label{sec:Introduction}}

\begin{figure}[htbp]
\begin{centering}\includegraphics[width=0.9\columnwidth,keepaspectratio]{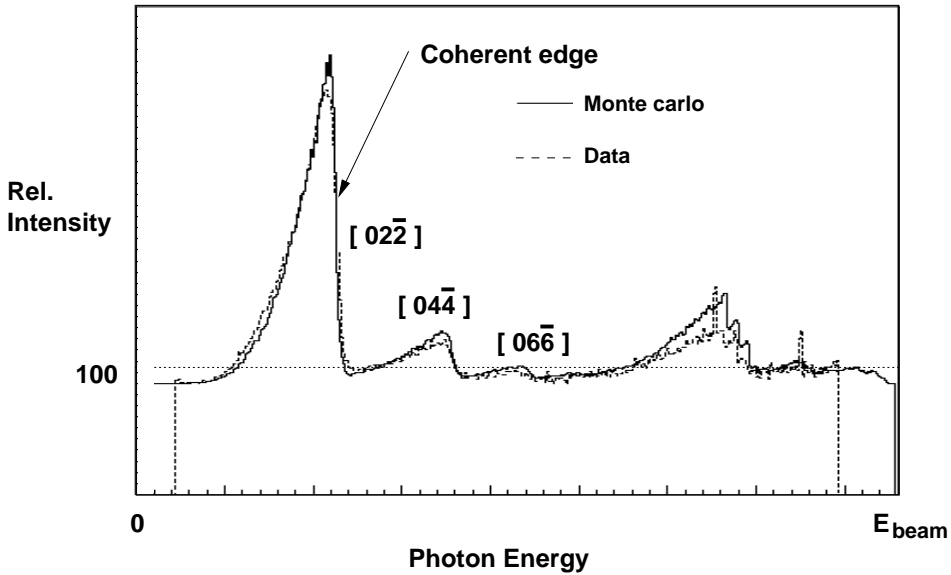}\par\end{centering}

\caption{Typical coherent bremsstrahlung enhancement spectrum. The spectrum
obtained from the crystal is divided though by a reference spectrum
from an amorphous radiator and normalised to be 100 in regions where
there is only a small coherent contribution. \label{cbrem_fig}}
\end{figure}

In the coherent bremsstrahlung technique a thin crystal oriented correctly
in an electron beam can produce photons with a high degree of linear
polarization \cite{1}. A typical photon energy spectrum is shown
in figure \ref{cbrem_fig}, where the region of high polarization
is under the peak to the left of the \emph{coherent edge} whose position
depends on the orientation of the crystal relative to the electron
beam. %
 The crystal is mounted on a goniometer to control its orientation,
but in order to be able to set up coherent bremsstrahlung it is first
necessary to measure an appropriate set of angular offsets between
the crystal axes and electron beam direction. A method for measuring
offsets and aligning the crystal was developed by Lohman et al, and
has been used successfully in Mainz \cite{2}. However, recent attempts
to investigate new crystals have shown that this approach has limitations
which become more serious at higher beam energies where more accurate
setting of the crystal angles, which scale with $1/E_{beam}$, is
required. (Eg. the recent installation of coherent bremsstrahlung
facility at Jlab, with $E_{beam}\simeq$ 6GeV ) This paper describes
a new alignment technique which overcomes these limitations. Wherever
possible, specific examples are given to illustrate the technique,
and section \ref{sec:A-practical-guide} outlines some general methods
for installing new crystals and measuring the angular offsets of the
electron beam in the goniometer reference frame.

\section{Coherent bremsstrahlung - a simplified view.\label{sec:Coherent-simple}}

\begin{figure}[htbp]
\begin{centering}\includegraphics[width=0.6\columnwidth,keepaspectratio]{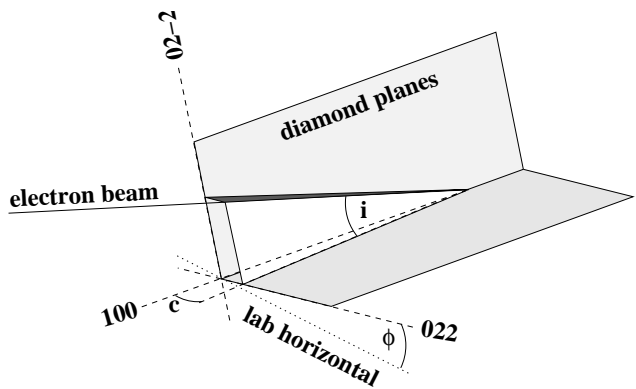}\par\end{centering}

\caption{Illustration of the scattering angles for coherent bremsstrahlung
from a diamond in the 100 orientation. The sets of planes defined
by the $[022]$ and $[02\overline{2}]$ lattice vectors are represented
by 2 single orthogonal planes. \label{simple_fig}}
\end{figure}

To provide a linearly polarised photon beam for an experiment we need
to be able to adjust the orientation of the polarization plane and
the shape of the photon intensity distribution which determines the
degree of linear polarization in the photon energy range of interest.
The main coherent peak is produced by scattering from one specific
set of crystal planes, represented by reciprocal lattice vector \textbf{\emph{g}}
= {[}$g_{x}g_{y}g_{z}]$. In practice a thin diamond crystal cut in
the 100 orientation is used and the main peak is produced by scattering
from the set of planes represented by the $[022]$ or $[02\overline{2}]$
reciprocal lattice vector, since this produces the highest polarization.
For simplicity, formulae and examples will be presented for these
conditions with some remarks on how to generalise to other crystals
in different orienations, and different lattice vectors. The parameters
of interest are illustrated in figure \ref{simple_fig}, where the
sets of crystal planes defined by the $[022]$ and $[02\overline{2}]$
lattice vectors are represented by 2 orthogonal surfaces.

The angle of the polarization plane is fixed by $\phi,$ the azimuthal
orientation of the 022 axis relative to the lab horizontal, and the
position of the main coherent edge (specifically, its fractional energy
$x=E_{edge}/E_{beam}$) is controlled by the angle \emph{c} between
the electron beam and the $[02\overline{2}]$ planes. In practice
then, \emph{c} must be adjusted to position the main coherent peak
at the region of interest. This will also set the position of subsequent
\emph{harmonics}. For example, in figure \ref{cbrem_fig} the main
peak is produced by scattering from the $[02\overline{2}]$ planes
and subsequent lower strength peaks from $[04\overline{4}]$ and $[06\overline{6}]$
are also clearly visible; all of these move \emph{in-concert} as the
angle $c$ between the beam and the $[02\overline{2}]$ planes is
adjusted. To control the remaining components of the distribution,
whilst keeping the main coherent peak (and harmonics) fixed, requires
adjustment of the angle $i$ between the electron beam and the direction
orthogonal to $c$. This angle will generally be set higher than $c$
by a factor of about 4, to position the coherent contributions from
the orthogonal planes at a high photon energy beyond the peak of interest. 

\begin{figure}[htbp]
\begin{centering}\includegraphics[width=0.3\columnwidth,keepaspectratio]{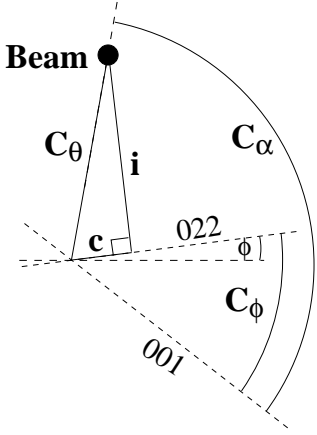}\par\end{centering}

\caption{Diagram showing the relationship between $c,i.\phi$ and crystal
angles $C_{\theta},C_{\alpha},C_{\phi}$. \label{ci_fig}}
\end{figure}

Referring to figure \ref{simple_fig}, we choose the values of $\phi,c,i$
as follows:

\begin{enumerate}
\item $\phi$ - the angle between the 022 axis and the horizontal.\\
\\
This is simply selected by deciding which orientation of the polarization
plane is best for the detector geometry of the experiment. Often,
for systematics, experiments are run with two orthogonal settings
of the polarization plane, $\phi$ and $\phi+\pi/2$. For practical
reasons this is most easily achieved by swapping the values of $c$
and $i$, rather than adjusting $\phi$. With $c,i$ as show in figure
\ref{simple_fig}, the polarization plane is $\phi$. Swapping them
makes the polarization plane $\phi+\pi/2$. A further common simplification
is to use PARA / PERP polarizations, where the polarization plane
is parallel or perpendicular to the lab horizontal.\\
 
\item $c$ - the small angle between the beam and the $[02\overline{2}]$
planes which fixes the position of the coherent edge.\\
\\
For a diamond in the 100 orientation, if we restrict our interest
to the position of the coherent edge from \textbf{\emph{g}} = $[022]$,$[02\overline{2}]$,$[044]$,$[04\overline{4}]$
etc. then $c$ (in radians) can be calculated as follows:''\begin{equation}
c\simeq\frac{k}{gE_{0}^{2}[\frac{1}{E}-\frac{1}{E_{0}}]}\label{eq:edge_angle}\end{equation}
where:\\
$g$= $\pm2,\pm4,\pm6$ ....\\
$E$ = required position of coherent edge (MeV) \\
$E_{0}$ = electron beam energy (MeV)\\
$k=m_{e}a/4\sqrt{2}\pi=$ 26.5601 MeV\\
$m_{e}=$ mass of electron = 0.511 MeV\\
$a$ = diamond lattice constant = 923.7 (dimensionless units)\\
\\
In practice, the coherent edge position $E$ should be reasonably
close to the calculated value, but this depends on how accurately
the initial angular offsets are measured during the setup (see section
\ref{sec:A-practical-guide}). The position of the coherent edge is
tuned by looking at enhancement spectra and making small adjustments
to $c$.\\

\item $i$ - the angle between the beam and the orthogonal planes. \\
\\
This is set to be larger than $c$ by about a factor of 4 and then
tuned using feedback from the enhancement spectra. It needs to be
adjusted to ensure that the main coherent peak is as \emph{clean}
as possible and that there are no interfering contributions from any
higher order lattice vectors.
\end{enumerate}
For the purposes of comparing with bremsstrahlung calculations it
is neccessary to describe the orientation in terms of the crystal
angles%
\footnote{In Lohman's paper \cite{2} the crystal angles are simple defined
as $\theta,\alpha,\phi$, but to make it clear that they relate to
the crystal coordinate system, as opposed to the goniometer system
they are labelled here as $C_{\theta},C_{\alpha},C_{\phi}$.%
} $C_{\theta},C_{\alpha},C_{\phi}$, where $C_{\theta},C_{\alpha}$
are the polar and azimuthal angles of the electron beam in the reference
frame of the crystal (defined by the 100, 010, 001 axes) and $C_{\phi}$
is the azimuth of \textbf{\emph{g}} in the same reference frame. This
is represented in figure \ref{ci_fig}, which is effectively a 2D
view of figure \ref{simple_fig} looking along the 100 axis. Here,
since the polar angles between the beam and crystal are small, $c$
and \textbf{\emph{$i$}} are represented by two lines parallel and
perpendicular to the 022 axis. This diagram shows that the relationship
between $\phi,c,i$ and $C_{\theta},C_{\alpha},C_{\phi}$ is rather
simple.

\begin{flushleft}\begin{equation}
C_{\phi}=\frac{\pi}{4}\label{eq:phi_c}\end{equation}
\begin{equation}
C_{\theta}=\sqrt{c^{2}+i^{2}}\label{eq:theta_c}\end{equation}
\begin{equation}
C_{\alpha}=C_{\phi}+\tan^{-1}\frac{i}{c}\label{eq:alpha_c}\end{equation}
\par\end{flushleft}

In summary, for a diamond in the 100 orientation with the coherent
peak coming from the $[022]$ or $[02\overline{2}]$ planes, we have
a simple prescription for deciding on values of $\phi,c,i$. We also
have simple set of equations (eq. \ref{eq:phi_c},\ref{eq:theta_c},\ref{eq:alpha_c})
for calculating the corresponding angles $C_{\theta},C_{\alpha},C_{\phi}$
in the crystal reference frame; this is important for comaprision
with bremsstrahlung calculations. What is also required is a fast,
reliable method of setting up and calibrating the goniometer to allow
the angles $c,i,\phi$ to be easily selected. A new method for doing
this is described in the following sections.

\section{Definition of angles and offsets\label{sec:Definition-of-angles}}

\begin{figure}[tbh]
\begin{centering}\includegraphics[width=0.8\columnwidth,keepaspectratio]{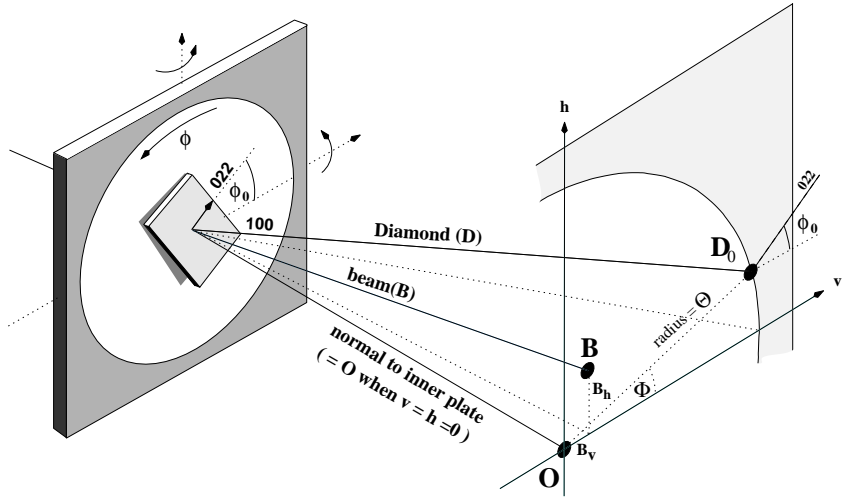}\par\end{centering}

\caption{Diamond radiator mounted on a goniometer showing the default angles
of the beam (\textbf{B}) and diamond (\textbf{D}), and definitions
of angular offsets. \label{goni}}
\end{figure}

As shown in Figure \ref{goni}, the diamond is attached to the inner
plate of the goniometer (white circle) which can rotate azimuthally
($\phi$) inside a nested pair of frames which can rotate about vertical
and horizontal axes ($v$ and $h$). The curved arrows on the figure
indicate the +ve direction of motion on each of the axes. The goniometer
should be set up in close alignment with the lab frame and the beam
direction. Explicitly, the settings on the $v$ and $h$ axes should
be zero (ie the origins defined in the goniometer setup) when these
axes are aligned with the lab vertical and horizontal, and the normal
to the inner plate \textbf{O} is aligned as closely as possible with
the beam direction \textbf{B}. Even with careful initial setup of
the goniometer there will be a small, but non-negligible, misalignment
with the beam direction \textbf{B}, which may also vary slightly each
time new beam is set up for an experiment. There will also be a misalignment
of the 100 diamond crystal axis \textbf{D} with the normal to the
inner plate \textbf{O} due to imperfections in the mounting and in
the cutting from the original stone. These angles are typically in
the range 0 - 60mrad, and are highly exaggerated in the diagram for
clarity. A simple and direct way of visualising the relative orientation
of the beam, goniometer and crystal is to project the vectors \textbf{B}
and \textbf{D} onto a 2D plane%
\footnote{A more accurate representation would use the surface of a sphere.
However, the angles involved are < 60 mrad, so the error introduced
by using a plane is negligible.%
}, where a +ve rotation about the $v$ axis moves \textbf{D} to the
right, and +ve rotation about the $h$ axis moves \textbf{D} upwards.
The origin of this system is defined by \textbf{O}, the orientation
of the normal to the goniometer in its zero position. The azimuthal
orientation of the 022 axis relative to the lab horizontal can also
be represented on this plot. The \textbf{D} vector sweeps out a cone
of angle $\Theta$ when the goniometer is rotated about it's azimuthal
axis. %

Figure \ref{goni} represents the crystal in its default position.
Ie. the goniometer has been set up and aligned approximately with
the beam direction, and a new crystal has been mounted with the 100
axis also approximately aligned with the beam direction. An appropriate
set of angular offsets to describe this are:

\begin{itemize}
\item $\phi_{0}$ - the azimuthal offset of the crystal as defined by the
angle between the 022 axis and the lab horizontal.
\item B$_{v}$,B$_{h}$ - the offsets between the beam direction \textbf{B}
and the origin \textbf{O}.
\item $\Theta,\Phi$ - the offsets of the 100 axis in the default position
\textbf{D}$_{0}$, in terms of its polar angular displacement $\Theta$
from the origin \textbf{O}, and the angle $\Phi$ between \textbf{D}$_{0}$
and the horizontal.
\end{itemize}
If these 5 offsets are known then it is possible to position the crystal
at any azimuthal angle, with any required small angular orientation
of the diamond \textbf{D} relative to the beam \textbf{B}. As described
in section \ref{sec:Coherent-simple}, the aim is to move the diamond
from its default orientation \textbf{D}$_{0}$ to some position \textbf{D}$_{\phi,c,i}$
where three angles $\phi,c,i$ are set at their required values. The
movement from \textbf{D}$_{0}$ to \textbf{D}$_{\phi,c,i}$ has to
be achieved by setting the appropriate angular coordinates on the
goniometer's azimuthal, vertical and horizontal axes $G_{a},G_{v},G_{h}$.
The derivation of these coordinates as functions of the angular offsets
and desired orientation angles ($\phi,c,i$) is illustrated in figure
\ref{cap:any-angle}, which shows a graph in angular displacement
about the \emph{v} and \emph{h} rotation axes. 

\noindent \begin{center}%
\begin{figure}[htbp]
\begin{centering}\includegraphics[width=0.6\columnwidth,keepaspectratio]{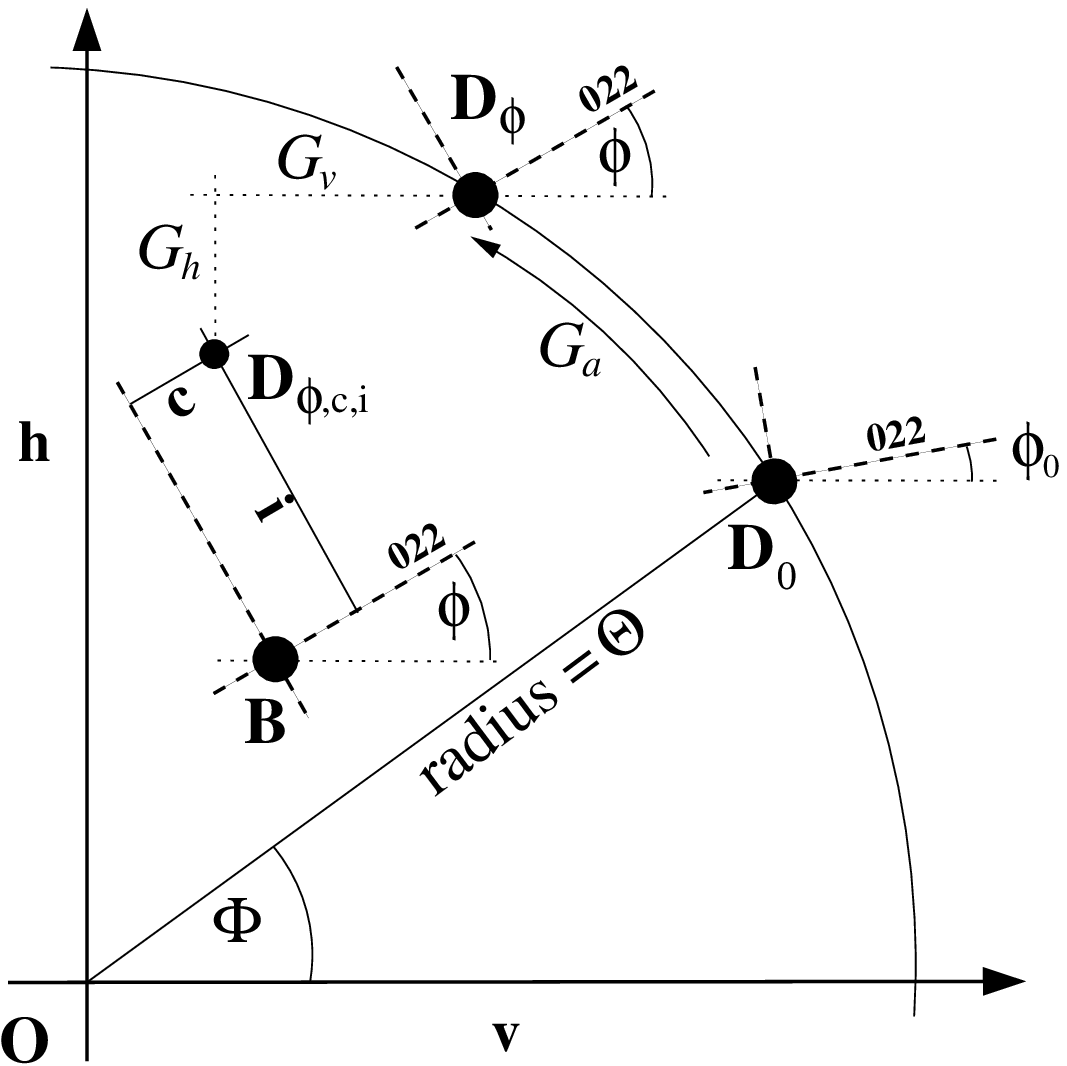}\par\end{centering}

\caption{\label{cap:any-angle}Angles involved in moving from default position
\textbf{D}$_{0}$ to an arbitrary position \textbf{D}$_{\phi,c,i}$. }
\end{figure}
\par\end{center}

The movement from \textbf{D}$_{0}$ to \textbf{D}$_{\phi,c,i}$ can
be thought of as a 2 step process: Firstly, the crystal needs to be
rotated azimuthally by adjusting $G_{a}$ to set the correct orientation
$\phi$ of the 022 axis. This has the effect of changing the angular
offset between the 100 diamond axis \textbf{D} and the beam \textbf{B},
shown as movement from \textbf{D}$_{0}$ to \textbf{D}$_{\phi}$ in
figure \ref{cap:any-angle}. Secondly, the goniometer's vertical and
horizontal rotation coordinates $G_{v},G_{h}$ must be adjusted to
set the correct angle between the diamond and the beam. In terms of
figure \ref{cap:any-angle} this means a translation from \textbf{D}$_{\phi}$
to \textbf{D}$_{\phi,c,i}$.

Applying simple trigonometry to figure \ref{cap:any-angle} gives
the following expressions for the goniometer settings $G_{a},G_{v},G_{h}$
required to achieve the diamond orientation \textbf{D}$_{\phi,c,i}$.

\begin{equation}
G_{a}=\phi-\phi_{0}\label{eq:phi_g}\end{equation}
 \begin{equation}
G_{v}=c\cos\phi-i\sin\phi-\Theta\cos(G_{a}+\Phi)+B_{v}\label{eq:theta_vg}\end{equation}
 \begin{equation}
G_{h}=c\sin\phi+i\cos\phi-\Theta\sin(G_{a}+\Phi)+B_{h}\label{eq:theta_hg}\end{equation}

Note, this will set the 022 axis at azimuthal angle $\phi$. However,
since the crystal has 4 fold symmetry about the 100 axis, interchanging
the values of $c$ and $i$ rotates the polarization plane by $\pi/2$,
and hence all orientations of the polarization plane are possible
with azimuthal rotations of $\leq\pi/4$ relative to the default orientation. 

This gives a prescription for selecting the orientation of the polarization
plane, and the energy of the coherent edge, provided the offsets $\phi_{0},\Phi,\Theta,B_{v},B_{h}$
are known. The challenge is to find a method of measuring these offsets
by scattering electrons from the diamond and interpreting shape of
the resulting photon energy spectra. The method described here is
the \emph{Stonehenge Technique}.

\section{Lohman's method and the need to extend it\label{sec:Scans}}

In the alignment technique developed by Lohman et al \cite{2}, angular
offsets are measured by carrying out a series of scans, each involving
a sequence of small angular movements (steps) of the crystal with
the corresponding accumulation of a photon energy spectrum (obtained
from a tagging spectrometer). This is divided through by an incoherent
spectrum from an amorphous radiator to highlight the coherent contribution
and eliminate the effect of channel to channel variations in the rates
on the individual counters of the tagging spectrometer. A single slice
of such a scan produces an enhancement plot such as that shown in
figure \ref{cbrem_fig}. The results of the complete scans are plotted
on 2d histograms which give information information about the orientation
of the crystal. This technique has been very successful, and the method
described here can be considered as an extension of it. In summary,
Lohman's technique relies on scans where only a single goniometer
coordinate is varied at once. The default azimuthal orientation $\phi_{0}$
is determined by scanning in $G_{a}$. The orientation of the crystal
022 vector is then selected to be $\phi=0^{\circ}$ to allow the orientation
of polarization plane to be $0^{\circ}$ or $90^{\circ}$(conventionally
PARA or PERP). Once the $G_{a}$ coordinate is fixed for PARA / PERP
conditions are considerably simplified since the $[022]$ and $[02\overline{2}]$
reciprocal lattice vectors are now aligned with the \emph{v} and \emph{h}
axes of the goniometer. Equations \ref{eq:theta_vg} and \ref{eq:theta_hg}
reduce to:\begin{equation}
G_{v}=c+k_{v}\label{eq:lohmanv}\end{equation}

\begin{equation}
G_{h}=i+k_{h}\label{eq:lohmanh}\end{equation}

Here, $k_{v}$ and $k_{h}$ are the angular offsets beween the beam
\textbf{B} and diamond \textbf{D} in this specific PARA / PERP orientation.
These offests can be simply determined with scans in $G_{v}$ and
$G_{h}$. Once $k_{v}$ and $k_{h}$ have been measured the coherent
peak can easily be put at any required energy setting for either PARA
or PERP polarization. For brevity only the essential concepts have
been described here, the full details are given in Lohman's paper
\cite{2}. 

Of all the offsets defined in figures \ref{goni} and \ref{cap:any-angle},
only $\phi_{0}$ is measured explicitly in Lohman's technique. There
is no need to measure the others if we adhere to the special circumstance
of only using the PARA / PERP orientations of the crystal. However,
setting up for PARA / PERP relies on measuring $\phi_{0}$, by scanning
in $G_{a}$. This is non-trivial, since any scan in $G_{a}$ makes
the crystal axis \textbf{D} describe a cone in space (see figure \ref{goni})
and therefore changes 3 parameters simultaneously ($\phi,c,i$). The
resulting 2D plot can be difficult (or impossible) to interpret. In
practice, the setting up of a new crystal is very time consuming,
and relies on the experience and intuition of the user, who has to
carry out a series of exploratory and iteratative scans until a consistent
picture is obtained. This can only be done reliably if the initial
misalignment between the beam \textbf{B}, goniometer \textbf{O} and
crystal \textbf{D} is small (ie $\leq$ $1/[2E_{beam}(GeV)]^{\circ}$).
A faster, more general technique is described here. This new method
is still based the interpretation of scans, but can cope with a larger
mounting misalignment and, since it measures all of the offsets described
in section \ref{sec:Definition-of-angles}, allows any arbitrary orientation
of the polarization plane to be selected.

\section{The Stonehenge Technique\label{sec:The-Stonehenge-Technique}}

\begin{figure}[htbp]
\begin{centering}\includegraphics[width=0.8\columnwidth,keepaspectratio]{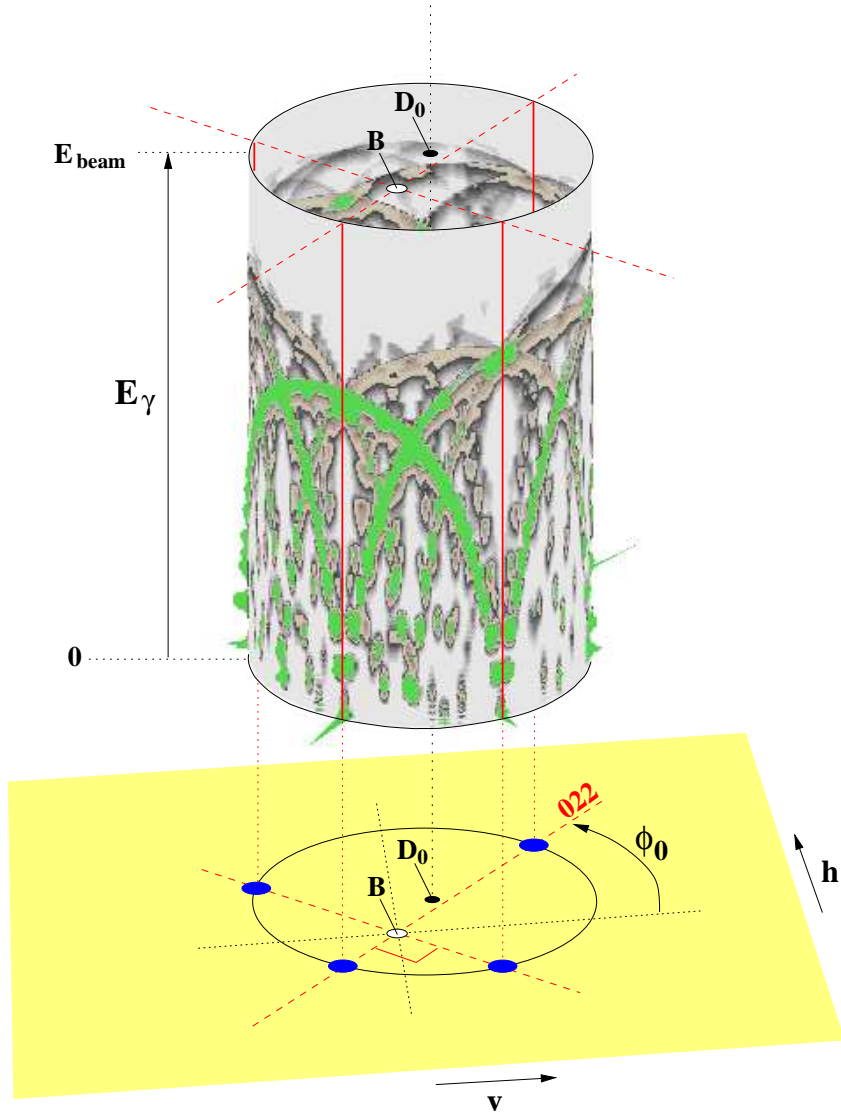}\par\end{centering}

\caption{\label{cyl.fig}Simulation showing the construction of a \emph{stonehenge
plot} from an \emph{hv-scan} on a crystal.}
\end{figure}

\begin{figure}[htbp]
\begin{centering}\includegraphics[width=0.8\columnwidth,keepaspectratio]{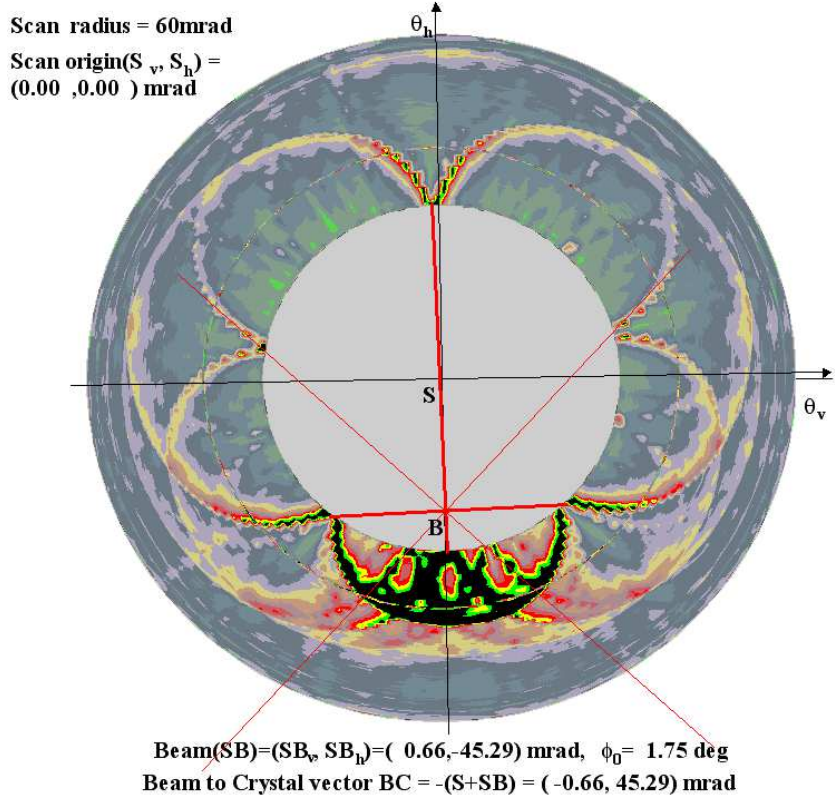}\par\end{centering}

\caption{\label{scan_phi_def.fig}An \emph{h}v scan and associated stonehenge
plot showing calculated offsets for the 100$\mu$m Mainz crystal in
the default orientation.}
\end{figure}

\begin{figure}[htbp]
\begin{centering}\includegraphics[width=0.8\columnwidth,keepaspectratio]{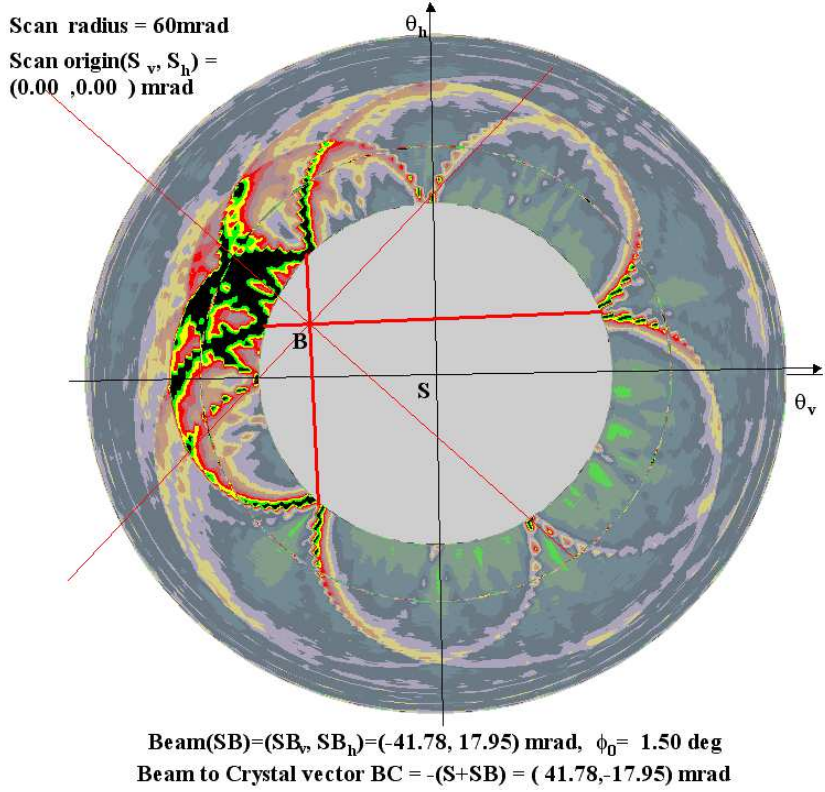}\par\end{centering}

\caption{\label{scan_phi_def180.fig}An \emph{hv} scan and associated stonehenge
plot showing calculated offsets for the Mainz 100$\mu$m crystal rotated
azimuthally by 180$^{\circ}$ from the default orientation.}
\end{figure}

The basis of the stonehenge technique is an \emph{hv} scan which sweeps
the crystal axis \textbf{D} around in a cone of angular radius $\theta_{r}$
by stepping sinusoidally on the $v$ and $h$ axes as follows: 

\begin{equation}
\begin{array}{c}
G_{v}=S_{v}+\theta_{r}\cos\left(\phi_{r}\right)\\
G_{h}=S_{h}+\theta_{r}\sin\left(\phi_{r}\right)\end{array}\left\} \begin{array}{c}
\end{array}0\leq\phi_{r}<2\pi\right.\label{eq:hv_scan}\end{equation}

where $S_{v},S_{h}$ are the coordinates of the center of the scan
relative to the zero positions on the goniometer axes. For each point
$G_{v},G_{h}$ in the scan an enhancement spectrum is measured, and,
since the intensity is a function of $E_{\gamma,}G_{v},G_{h}$, a
full representation of the data can be made by plotting it on surface
of a cylinder (figure \ref{cyl.fig}). %
Regions of high intensity form sets of curves which show the coherent
contributions from different sets of crystal planes as their angles
change relative to the beam. The curves with the strongest intensity
relate to scattering from the sets of planes defined by the $[022]$
and $[02\overline{2}]$ reciprocal lattice vectors, and the points
in the scan where they converge at $E_{\gamma}=0$ indicate where
the relevant set of planes is parallel to the beam. Vertical lines
are drawn though these 4 points in figure \ref{cyl.fig} and projected
onto the perimeter of a circle to produce a \emph{stonehenge plot}.
By fitting a pair of orthogonal lines to these 4 points the beam position
can be identified. The azimuthal offset $\phi_{0}$ of the crystal
can be measured, and the offset vector $\overrightarrow{BD_{0}}$
between the crystal axis default position \textbf{D}$_{0}$ and the
beam direction \textbf{B} determined. Notice that the choice of which
of the 2 lines represents the 022 axis is somewhat arbitrary due to
the the 4-fold symmetry, so we choose such that $\phi_{0}$ is between
0 and $\pi/2$. For practical reasons the cylinder shown in figure
\ref{cyl.fig} is flattened out onto the outer part a polar diagram
with photon energy increasing in the outward radial direction %
\footnote{In the initial development, the offsets were obtained by analyzing
only the high intensity spots on the perimiter of a circle. This is
why the \emph{Stonehenge Technique} is so called. %
}. The stonehenge plot is then constructed in the center. Figures \ref{scan_phi_def.fig}
and \ref{scan_phi_def180.fig} show the result of such measurements
carried out in Mainz using a 100$\mu$m crystal. The 100 axis was
made to describe a cone in 180 x 2$^{\circ}$ steps by moving the
goniometer axes as described by eqn. \ref{eq:hv_scan}, with $\theta_{r}=60$
mrad and $S_{v}=S_{h}=0.$%

The plotting and analysis was carried out using a ROOT macro \cite{3,4}
which was used to superimpose lines on the polar plot on the basis
of 4 selected points (corresponding to the $[022]$ and $[02\overline{2}]$
vectors). This was carried out repeatedly until a pair of orthogonal
lines was found to give a good fit to the 4 points (judged by eye).
The intersection of these lines defines the coordinates of the beam
relative to the center of the scan. Further corroboration that the
correct points have been selected is given by the lines at 45$^{\circ}$
to the main pair, which pass though the points related to the {[}004]
and {[}040] vectors. The crystal's azimuthal orientation $\phi_{0}$
is found from the gradient and the vector $\overrightarrow{BD_{0}}$
calculated as follows: $\overrightarrow{BD_{0}}=-(\overrightarrow{S}+\overrightarrow{SB})$,
where $\overrightarrow{S}$ ($S_{v},S_{h}$) is the origin of the
center of the \emph{hv} scan in relation to the goniometer origin
\textbf{O}. Initially $\overrightarrow{S}$ will be (0,0) but if extra
precision is required the scan can be repeated with $\overrightarrow{S}$
set to the first measured value of $\overrightarrow{SB}$ and a smaller
radius $\theta_{r}$. Figures \ref{scan_phi_def.fig} and \ref{scan_phi_def180.fig}
show the output of the ROOT macro with the template correctly fitted
and all the values calculated. 

\begin{figure}
\begin{centering}\includegraphics[width=0.8\columnwidth,keepaspectratio]{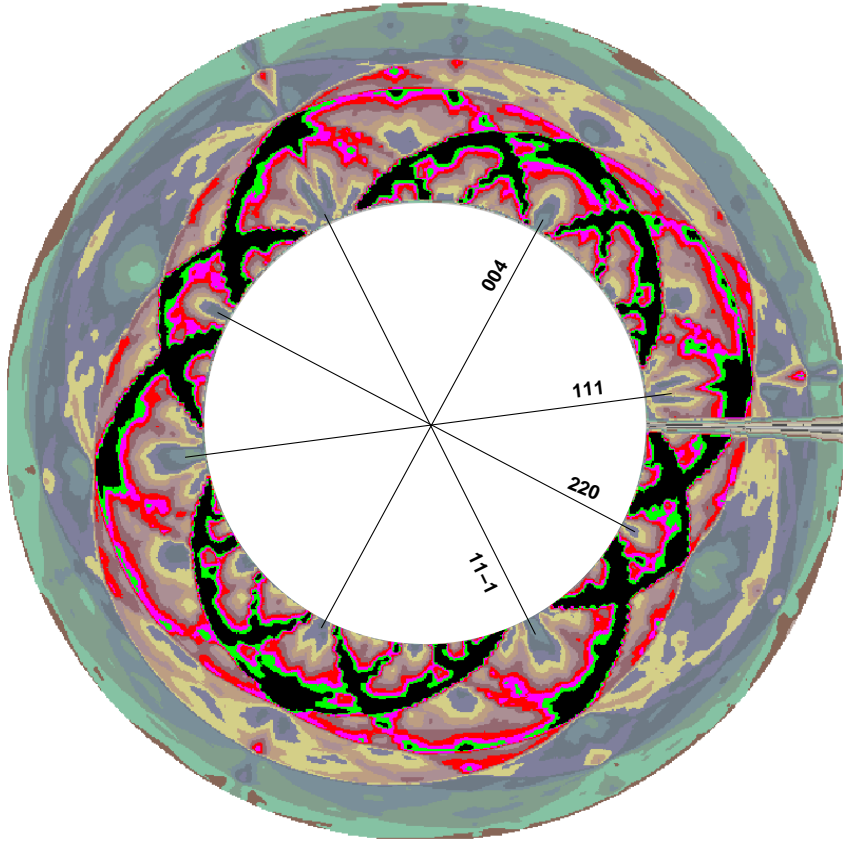}\par\end{centering}

\caption{Stonehenge plot of a diamond in the 110 orientation. Here, the crystal
was aligned with the beam and the 111 axis set at 10$^{\circ}$ to
the horizontal for measurements of channeling radiation. \label{fig:110plot}}
\end{figure}

Initially the Stonehenge technique seems rather abstract, so it is
worth pointing out some of its advantages. A single $hv$-scan with
corresponding Stonehenge plot allows the simultaneous measurement
of the azimuthal position of the crystal and the vertical and horizontal
angles it subtends relative to the beam. With the previous method
each of these had to be measured separately through a long process
of iterations, with the caveat that the mounting misalignment had
to be small. Here, even in conditions where there is a large mounting
misalignmet, the scan can be repeated with a large enough radius to
produce an unambiguous plot. Also, if only PARA / PERP running conditions
are required, the azimuthal orientation is set to $0^{\circ}$ by
adjusting $G_{a}$ and the scan repeated to find the offsets $k_{v},k_{h}$
from eq. \ref{eq:lohmanv},\ref{eq:lohmanh} (more details are given
in section \ref{sec:A-practical-guide}). Furthermore, the speed and
reliability of the Stonhenge technique make it practical to measure
all of the offsets defined in section \ref{sec:Definition-of-angles}
and really allows the choice of any arbitrary values of $\phi,c,i$
(see following section). Finally, it is relatively simple to align
a crystal with a different structure, or in a different orientation;
we just need to calculate which lattice vectors will give the strongest
contribution to the coherent bremsstrahlung and overlay a corresponding
template on the Stonehenge plot. For example, figure \ref{fig:110plot}
shows the data from a diamond in the 110 orientation after alignment,
where the template consists of lines in the 11$\pm1$, 220 and 004
directions. %

\section{Obtaining a complete set of offsets from scans\label{sec:Obtaining-offsets-from}}

As shown in the previous section, it is possible to measure the azimuthal
orientation of the crystal and the offset between the beam \textbf{B}
and crystal axis \textbf{D} with a single scan. For many experiments
this may provide enough information to run at all the required conditions
- eg. only in PARA / PERP mode. However, we can achieve more generality.

\begin{figure}[htbp]
\begin{centering}\includegraphics[width=0.5\columnwidth,keepaspectratio]{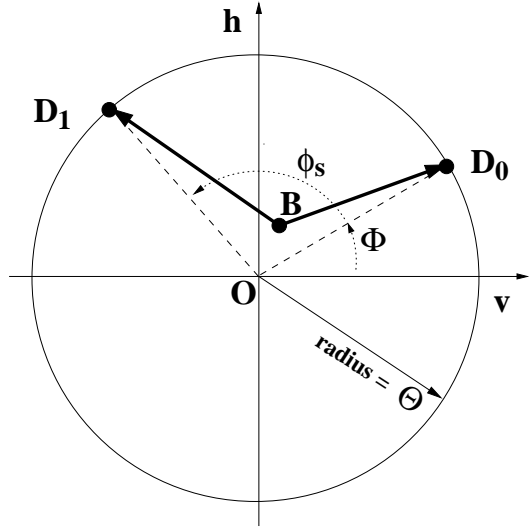}\par\end{centering}

\caption{Illustration of how to determine derive the offsets $\Phi,\Theta$
and the beam position $B$ from the values of $\protect\overrightarrow{BD_{0}}$
and $\protect\overrightarrow{BD_{1}}$ obtained from two $hv$ scans
carried out with an azimuthal separation of $\phi_{s}$. \label{theory.fig}}
\end{figure}

\begin{figure}[htbp]
\begin{centering}\includegraphics[bb=100bp 100bp 450bp 450bp,clip,width=0.5\columnwidth,keepaspectratio]{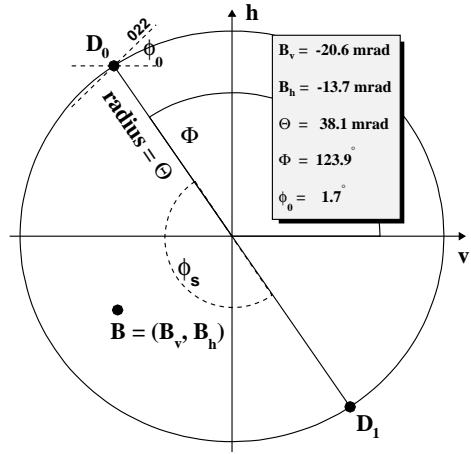}\par\end{centering}

\caption{\label{origin_data.fig}Calculation of the crystal offsets from the
results of two \emph{hv} scans on the Mainz {[}100]$\mu$m diamond.}
\end{figure}

We represent the offset between the beam \textbf{B} and crystal axis
\textbf{D} with a vector $\overrightarrow{BD_{0}}$, measured from
an initial scan. If the goniometer is rotated azimuthally by some
known amount $\phi_{s}$, another scan can be carried out to determine
a different offset $\overrightarrow{BD_{1}}$ . The vectors $\overrightarrow{BD_{0}}$,
$\overrightarrow{BD_{1}}$ with a known angular separation $\phi_{s}$($=\angle D_{0}OD_{1}$)
provide the information to locate the origin, construct figure \ref{cap:any-angle}
and yield the required offsets. This relationship between $O,\overrightarrow{BD_{0}},\overrightarrow{BD_{1}},\Theta,\Phi$and
$\phi_{s}$is shown in figure \ref{theory.fig}. %
If the vector components in $v,h$ are $\overrightarrow{BD_{0}}=(v_{0},h_{0})$,
$\overrightarrow{BD_{1}}=(v_{1},h_{1})$ then the offsets $B_{v},B_{h},\Theta,\Phi$
are calculated as follows:

\begin{equation}
B_{v}=-\left[\frac{(v_{0}+v_{1})}{2}-\frac{(h_{1}-h_{0})}{2\tan(\phi_{s}/2)}\right]\label{eq:theta_vb}\end{equation}
\begin{equation}
B_{h}=-\left[\frac{(h_{0}+h_{1})}{2}+\frac{(v_{1}-v_{0})}{2\tan(\phi_{s}/2)}\right]\label{eq:theta_hb}\end{equation}
\begin{equation}
\Phi=\arctan\left[\frac{B_{h}+h_{0}}{B_{v}+v_{0}}\right]\label{eq:phi_t}\end{equation}
\begin{equation}
\Theta=\sqrt{(B_{v}+v_{0})^{2}+(B_{h}+h_{0})^{2}}\label{eq:theta_t}\end{equation}
 In practice, this means carrying out an initial scan to determine
$\overrightarrow{BD_{0}}$, rotating by an angle $\phi_{s}$ on the
goniometer's azimuthal axis, and repeating the scan to determine $\overrightarrow{BD_{1}}$.
Figures \ref{scan_phi_def.fig} and \ref{scan_phi_def180.fig} show
the results of such a pair of scans, where the crystal was rotated
azimuthally by $\phi_{s}=180^{\circ}$ between scans. As would be
expected, the position of the crystal relative to the beam has changed
considerably, but both scans agree (to within 0.5$^{\circ}$) on the
azimuthal orientation of the crystal. The components of $\overrightarrow{BD_{0}}$
and $\overrightarrow{BD_{1}}$ obtained from the two corresponding
stonehenge plots are inserted into eqns \ref{eq:theta_vb}-\ref{eq:theta_t}
to calculate the offsets and construct specific version of the diagram
shown in figure \ref{theory.fig}. The result obtained from the analysis
of the 2 example plots (figures \ref{scan_phi_def.fig} and \ref{scan_phi_def180.fig})
is shown in figure \ref{origin_data.fig}, which was generated from
a ROOT macro \cite{3,4}.

\begin{figure}[htbp]
\begin{centering}\includegraphics[width=0.8\columnwidth,keepaspectratio]{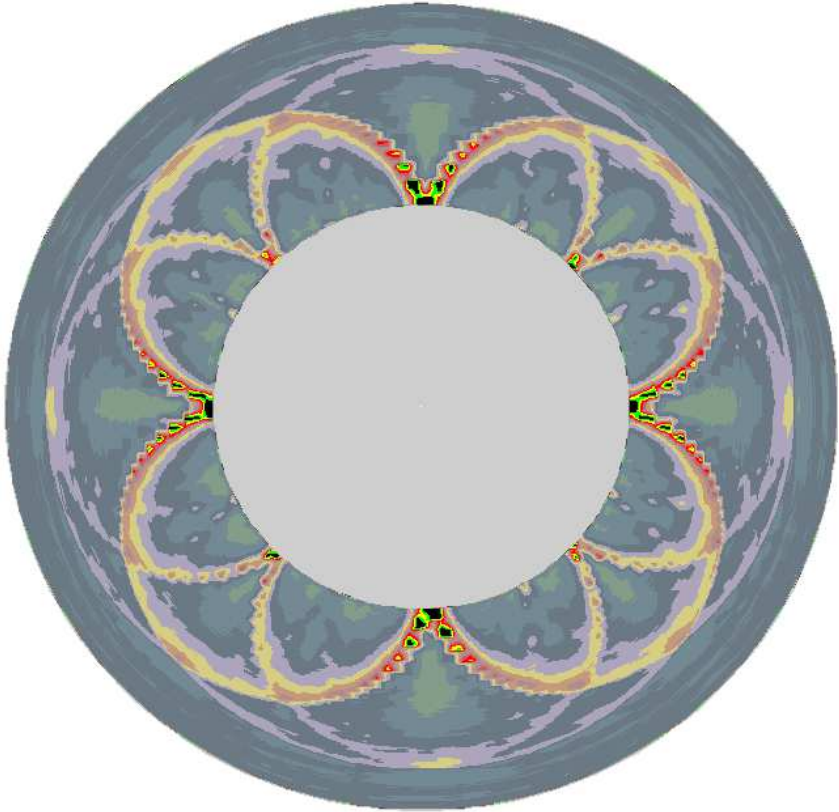}\par\end{centering}

\caption{\label{scan_perfect.fig}An hv scan of the Mainz $\mu$m crystal
in the $\phi_{d}=0$ orientation with all offsets installed (ie center
aligned perfectly with the beam). }
\end{figure}

With the offsets measured, eqns \ref{eq:phi_g} - \ref{eq:theta_hg}
can be to used to calculate the goniometer settings required to position
the crystal at any desired azimuthal angle ($\phi$) and any desired
orientation of the $[022]$ and $[02\overline{2}]$ vectors relative
to the beam (ie any values of \textbf{\emph{c}} and \textbf{i}). To
confirm this, we can attempt to set $\phi=0^{\circ}$ , align the
crystal perfectly with the beam and carry out another $hv$ scan,
which should have symmetry about the $v$ and $h$ axes. An example
is shown in figure \ref{scan_perfect.fig}.

The Stonehenge Technique has now been used successfully at MAMI in
Mainz, CB-ELSA in Bonn, and CLAS at Jefferson Laboratory to align
several diamonds. Based on experience at these facilities, the final
section is practical, step by step guide on how to setup a linearly
polarised photon beam using a combination of the Stonehenge Technique
and the methods described by Lohman \emph{et al.} \cite{2}.

\section{A step by step guide.\label{sec:A-practical-guide}}

The equations and example plots presented in the earlier sections
of this article give a prescription for aligning the crystal relative
to the beam for the most general circumstances; ie it is assumed that
none of the offsets ($\phi_{0},\Phi,\Theta,B_{v},B_{h}$) are known,
and that it is desirable to be able to select any arbitrary orientation
of the polarization plane. In practice, this full generality is seldom
required. For example, the crystal may have already been installed
for an earlier running period, or only the PARA / PERP orientations
of the polarization plane may be required. Hence, the steps described
below give a suggested procedure for the initial installation and
setup of the goniometer, subsequent alignment of new crystals and
re-calibration prior to each new run period, and some shortcuts which
can be taken for certain specific conditions. It is assumed here that
there is a combination of software and hardware in place to move the
goniometer in small angular increments and acquire the corresponding
photon energy spectra (ie carry out the scans). Several ROOT macros
\cite{3} have been developed to carry out the analysis of scans and
calculate the goniometer settings. These are available on the web
\cite{4}, or by sending an email to the author.

\subsection*{Collimation - before and after.}

The bremsstrahlung photon beam is tightly collimated to increase the
enhancement and the degree of polarization. Ideally it should be collimated
to less that half the characteristic angle of the bremsstrahlung cone
(eg. 2mm diameter at 23m at CLAS where $E_{beam}\sim5GeV$). For a
true picture of the collimated beam on target, the enhancement needs
to be derived from event by event data, where the DAQ is triggered
by the photon, and the energy derived by looking at the electron which
makes a timing coincidence with the tagger focal plane. Even at a
DAQ rate of $\sim5kHz$ it takes of the order of 30min to build up
enough statistics for a reasonable enhancement spectrum. This is certainly
useful for measuring the degree of polarization of the beam, but is
not adquate for anything requiring fast feedback, such as setup scans,
or steering the beam throught the collimator.

The scans described in this paper are based on scalers attached to
the tagger focal plane. These count at a high rate (eg $\sim10^{5}Hz$
per counter) and show the distributions of the scattered bremsstrahlung
electrons. They give a very fast enhancement spectrum for the puposes
of setup scans, and for setting the coherent edge to the correct position
in the photon energy spectrum. The can also provide a fast online
monitor that the coherent peak is stable. However, this scaler-derived
information does not tell us whether any of the photons made it through
the collimator to the experiment. 

The distinction between this scaler-based enhancement and a DAQ-based
enhancement is important. Just because the coherent peak is in the
correct place doesn't mean the polarized photons are getting through
the collimator to the experiment, and there must be adequate other
diagnostics in place to ensure that this happens.

\subsection*{Setup of the goniometer}

\begin{enumerate}
\item The positive directions of rotation of the goniometer should be defined
in software to correspond to the directions indicated in figure \ref{goni},
and the settings on the $v$ and $h$ axes should be zero (ie define
$G_{v}=G_{h}=0$) when these axes are aligned with the lab vertical
and horizontal, and hence the normal to the inner plate is aligned
as closely as possible with the beam direction (generally using a
laser and a mirror mounted flat on the plate). 
\item Despite the best efforts to setup the goniometer, it is likely that
there will still be non-negligible offsets ($B_{v},B_{h}$) between
the beam direction and the goniometer. If the beam position is well
defined by demanding that it pass through a small collimator, these
offsets can be measured during the installation of the first crystal
(see below) and used to redefine the zero positions (ie redefine $G_{v}=G_{h}=0$
on the $v$ and $h$ axes of the goniometer. This will mean give a
better default alignment between the beam and goniometer.
\end{enumerate}

\subsection*{Reference spectrum}

As with the method described by Lohman et al \cite{2}, all photon
energy spectra obtained with the crystal should be divided though
by a reference spectrum obtained with an amorphous radiator, which
should be taken immediately prior to attempting to align any crystals.
This divides out most of the incoherent bremsstrahlung contribution
to the spectrum and highlights the coherent peaks. It also smooths
out spikes which are due to the differences in efficiency of the counters
to produce a relatively smooth \emph{enhancement} spectrum. In a highly
segmented counter there are likely to be several counters which are
dead, or have very low efficiency, and it is recommended that these
channels are flagged by entering zeros as appropriate in the reference
spectrum. It is then a simple matter to ignore them in the division
part of the software and copy the the value from the adjacent (or
nearest good) channel. To compare successive steps in the scans it
is also necessary to normalize the divided spectra to the same baseline
(eg.$\sim100$). This should be done by multiplying the value in all
channels by a normalization factor of {[}100] $\div$ X, where X is
the value of the lowest non-zero channel in the divided spectrum (ie
corresponding to a point where there is the smallest possible coherent
contribution). This channel will in general be different, and has
to be recalculated, for each step in the scan.

\subsection*{Setting up the first crystal}

This describes the full alignment procedure on the assumption that
all offsets are to be measured, and that is is desireable to be able
to \emph{dial up} any reasonable set of $\phi,c,i$ values. In general
only some subset of these steps will be required depending on the
required running conditions and on information from any previous alignment.

\begin{enumerate}
\item The crystal should be mounted with its face parallel to the goniometer's
inner plate. Experience has shown that is reasonable to expect to
mount this to better than 2$^{\circ}$ (35mrad).
\item An initial \emph{hv} scan should be carried out (eqn \ref{eq:hv_scan})
with a radius of $\theta_{r}$ = 60mrad, using 180$\times2^{\circ}$
steps in $\phi_{r}$, and setting $S_{v}=S_{h}=0$. The resulting
data should be used to construct a stonehenge plot as described in
section \ref{sec:The-Stonehenge-Technique}, which should give the
default azimuthal orientation of the crystal $\phi_{0}$ and the offset
between the {[}100] axis and the beam direction $\overrightarrow{BD_{0}}$.
If there is ambiguity in this plot, it is likely to be because the
offset between the crystal and the beam is greater than 60mrad and
the beam spot lies outside the circle of the stonehenge plot. The
scan should be repeated with a greater value of $\theta_{r}$, eg.
120mrad. Once an initial value for $\overrightarrow{BD_{0}}$ has
been obtained, it is useful to \emph{zoom in} and repeat the scan
to obtain more accurate value of $\overrightarrow{BD_{0}}$. This
is done by setting $\vec{S}=-\overrightarrow{BD_{0}}$ (ie $S_{v}=-v_{0},S_{h}=-h_{0}$)
and redoing the original scan with smaller value of $\theta_{r}$.
In most situations an initial scan at $\theta_{r}$ = 60mrad, followed
by a second scan and $\theta_{r}$ = 20mrad is adequate. 
\item The goniometer should be rotated by a known amount $\phi_{s}$ (ideally
= 180$^{\circ}$ if the goniometer and cabling allows), and step 2
repeated. The azimuthal angle measured from the stonehenge plot should
be $\phi_{0}$+ $\phi_{s}$, and should yield the same value of $\phi_{0}$
as measured in step 2, to within about 0.5$^{\circ}$. The second
offset between the {[}100] axis and the beam direction $\overrightarrow{BD_{1}}$
should also obtained from plot.
\item The values of $\overrightarrow{BD_{0}}$ and $\overrightarrow{BD_{1}}$
and $\phi_{s}$ should be used to construct a plot like the one shown
in figure \ref{origin_data.fig} and calculate all the offsets ($\phi_{0},\Phi,\Theta,B_{v},B_{h}$)
\item The values of $\phi_{0},\Phi,\Theta$ are specific to this installation
of the crystal and should not change unless the crystal is remounted.
They should be recorded.
\item The values $B_{v},B_{h}$ describe the offset of the beam. At this
point it is sensible redefine the origin of the goniometer to coincide
with the beam. Ie. move the goniometer to $G_{v}=B_{v}$ and $G_{h}=B_{h}$
and redefine the zero positions in the goniometer software at these
positions. The default values of the beam offsets can now be set to
zero ($B_{v}=B_{h}=0$) and should only change if the beam position
changes or the goniometer is re-installed.
\item To confirm that the offsets have been correctly measured a final $hv$
scan should be carried out with the crystal in the required azimuthal
setting (say $\phi=0$), and the center of rotation aligned with the
beam. Setting $\phi=i=c=0$ in eqns \ref{eq:phi_g}-\ref{eq:theta_hg}
gives the coordinates $G_{\phi},G_{v},G_{h}$ for the center of scan
($S_{v},S_{h})$.
\end{enumerate}

\subsection*{Installation of other crystals}

Other crystals should be aligned in a similar way to the first crystal
but with step 6 omitted since the goniometer's zero setting should
now be consistent with the beam direction. Scans on new crystals should
confirm the beam position, which should be very close to the origin.

\subsection*{Setting the polarization plane }

Since there is 4-fold rotational symmetry about the azimuthal axis,
the goniometer need never be rotated by more than $\pm45^{\circ}$
to position the crystal lattice for any required orientation of the
polarization plane. For example, if we select $\phi=30^{\circ}$ (the
azimuthal angle of the $022$ crystal axis from the horizontal) and
select $c$ and $i$ to be the scattering angles from the $02\overline{2}$
and $022$ axes respectively, as shown in figure \ref{goni}c, then
the polarization plane with be at 30$^{\circ}$. Interchanging $c$
and $i$ will rotate the plane by 90$^{\circ}$ to 120$^{\circ}$
(or -60$^{\circ}$). Eqn \ref{eq:phi_g} should be used to find the
goniometer value $G_{\phi}$ to set the correct azimuthal orientation
of the crystal.

\subsection*{Calibrating the crystal. }

Once the required azimuthal angle $\phi$ has been set and the required
orientation of the polarization plane chosen ($\phi_{pol}=\phi$ or
$\phi_{pol}=\phi\pm90^{\circ}$ ), an energy calibration scan should
be carried out to produce a table of goniometer angles vs. coherent
edge. This involves fixing the incoherent scattering angle $i$ and
stepping through a range of coherent angles $c$. For example, in
Mainz where $E_{beam}=855MeV$, the incoherent scattering angle $i$
is set to 60mrad and a typical energy scan would be from -1mrad to
+10mrad in steps of 0.25mrad in $c$. In the simplest situation, where
the polarization plane is vertical or horizontal, it requires motion
on only a single axis, $v$ or $h$, and is described in Lohman's
paper \cite{2}. However, in the more general situation keeping $i$
fixed and stepping in $c$ requires movements on both the $v$ and
$h$ axes. Here eqns \ref{eq:phi_g}-\ref{eq:theta_hg} can be used
to generate a table of the goniometer settings $G_{v},G_{h}$ corresponding
to each $c$ value in the scan. The goniometer settings $G_{v},G_{h}$
required to set the position of the coherent edge to the required
photon energy are found by interpolating between the closest points
in the table. It is worth stressing here that this calibration is
appropriate only for the selected value of the polarization plane
$\phi_{pol}$. If this changes, then a new calibration run must be
made.

\subsection*{Re-tuning and re-calibrating}

If the position of the coherent peak drifts by a small amount during
running (eg. due to drift in the beam position) it can be compensated
for by making small adjustments to $c$, which in general means adjusting
the values to both the $G_{v}$, $G_{h}$ settings, again by interpolation
in the calibration table. If there is a large drift in the coherent
peak position then it is worth carrying out an \emph{hv} scan with
$\phi$ remaining in its current position and the center $S_{v},S_{h}$
corresponding to $c=i=0$ (see eqns \ref{eq:theta_vg},\ref{eq:theta_hg}).
The resulting stonehenge plot should confirm the value of the current
azimuthal orientation $\phi$, but should be give a non-zero value
for $\overrightarrow{SB}$; ie beam position should be offset from
the center of the scan. Note that on this information alone it is
not possible to say whether the crystal has moved in some way (eg.
a from expansion of a mounting wire) or the beam position has changed.
However, provided we want to keep the current azimuthal orientation
of the crystal $\phi$, this is not important. The components of $\overrightarrow{SB}$
(labeled $SB_{v},SB_{h}$ in figures \ref{scan_phi_def.fig} and \ref{scan_phi_def180.fig})
can be added to the current beam offsets $B_{v},B_{h}$ and the calibration
scan described above can be repeated.

\section*{Shortcuts}

\subsubsection*{A single azimuthal orientation.}

In many experimental situations it is only necessary to select a single
azimuthal orientation $\phi$ of the crystal and to set the polarization
plane. This will then be used for one specific orientation of the
polarization plane, or for the pair of orthogonal settings which are
parallel and perpendicular to this direction ( ie setting $\phi_{para}=\phi$
and $\phi_{perp}=\phi+90^{\circ}$ and interchanging the coherent
and incoherent angles $c,i$ ). Here, once the default azimuthal orientation
of the crystal $\phi_{0}$ has been determined, the desired azimuthal
orientation $\phi$ can be set, and the offset between the beam \textbf{B}
and crystal axis \textbf{D} measured specifically for this value of
$\phi$. This is achieved by carrying out a procedure similar to that
outlined as the first step for installing the first crystal. However,
as soon as the default azimuthal orientation $\phi_{0}$ of the crystal
is found with adequate precision, the goniometer setting $G_{a}$
should be adjusted to set the crystal to the desired azimuthal orientation
($G_{a}=\phi-\phi_{0}$). The \emph{hv} scan should then be repeated
in this new azimuthal orientation and the value of $\overrightarrow{BD_{0}}(v_{0},h_{0})$
found using the stonehenge plot. The beam offsets are now $B_{v}=-v_{0}$
and $B_{h}=-h_{0}$, and eqns \ref{eq:theta_vg} and \ref{eq:theta_hg}
simplified as follows:\begin{equation}
G_{v}=c\cos\phi-i\sin\phi+B_{v}\label{eq:theta_vg_simp1}\end{equation}
 \begin{equation}
G_{h}=c\sin\phi+i\cos\phi+B_{h}\label{eq:theta_hg_simp1}\end{equation}

The crystal can now be calibrated as described above, this time using
eqns \ref{eq:theta_vg_simp1} and \ref{eq:theta_hg_simp1} to generate
a series of goniometer settings for a scan in $c$ at a polarization
plane of $\phi_{pol}=\phi$. Again, interchanging $c$ and $i$ in
eqns \ref{eq:theta_vg_simp1} and \ref{eq:theta_hg_simp1} will give
the appropriate values for the $\phi_{pol}=\phi+90^{\circ}$.

\subsubsection*{PARA and PERP}

Notice, that if the PARA / PERP settings are being used (ie $\phi=0^{\circ}$
or $\phi=90^{\circ}$) eqns \ref{eq:theta_vg_simp1} and \ref{eq:theta_hg_simp1}
become simplified further and $c,i$ are related independently to
$G_{_{V}},G_{h}$.

\section{Conclusions}

A new method of aligning crystals for coherent bremstrahlung facilities
has been described. As with previous methods it still based the interpretation
of scans, but can cope with a relatively large mounting misalignment
and allows any arbitrary orientation of the polarization plane to
be selected. The technique has now become the standard method used
for setting up coherent bremsstrahlung at the several of world's main
coherent bremsstrahlung facilities (MAMI at Mainz, CLAS at Jefferson
Lab, ELSA at Bonn and MAXLab at Lunz).

\begin{ack}
The technique described in this paper were developed using data taken
at MAMI microtron facility in Mainz, using the Glasgow tagger.  Thanks
to all the members of the Mainz A2 collaboration who participated,
and in particular to Axel Schmidt and Roman Leukel from Mainz who
shared their knowledge and expertise on coherent bremsstrahlung. Thanks
also to Bob Owens, Cameron McGeorge and Jim Kellie from the Glasgow
group, who provided valuable feedback on the development of the technique
and production of this paper. All coherent bremsstrahlung calculations
to simulate experimental data were carried carried out with a modified
version of the ANB (ANalytic Bremsstrahlung) code provided by Alex
Natter from the Tuebingen group\cite{5}. 
\end{ack}
The work presented here was funded by the Engineering and Physical
Sciences Research Council (EPSRC) and by Eurotag, a Joint Research
Activity within the Eurpean Framework 6 I3HP initiative.

\end{document}